# Decentralization and Governance in IoT

## *Bitcoin and Wikipedia Case*

Anass Sedrati
*Mathematics, Computer& Networking Department*
*INPT*
Rabat, Morocco

RISE – Research Institutes of Sweden

Nelly Stoyanova
*Information Society Policy Department – Ministry of Transport, IT and Communications*
Sofia, Bulgaria

Abdellatif Mezrioui
*Mathematics, Computer& Networking Department*
*INPT*
Rabat, Morocco

Abdelaziz Hilali
*Mathematics, Computer& Networking Department*
*INPT*
Rabat, Morocco

Aziza Benomar
*Mathematics, Computer& Networking Department*
*INPT*
Rabat, Morocco

***Abstract*— In the era of digital revolution many contemporary events that changed the world were shaped through the Internet. Nowadays, the emergence of Internet of Things, combining physical objects with virtual networks, is expected to have even more influence. This new "decentralized" structure in the world raises questions such as power, governance and the notion of democracy online. The aim of this paper is to investigate these notions. We have taken the examples of Bitcoin and Wikipedia and examined their decision-making process. Our analysis has found some inconsistencies in their policies that are in contradiction with democracy and consensus principles of governance. Starting from our findings, we present further improvements that can be used to achieve more democracy and equity in the digital context.**

### I. INTRODUCTION (*HEADING 1*)

Described as the technology of the future, Internet of Things (IoT) will allow "objects" to be connected to different networks and become "smart". In some configurations, these objects can be directly connected and communicate with each other without human intervention. One of the challenges following this development is the governance of decentralized IoT systems in terms of power relations and decision-making. How democratic will be the electronic systems that we will use in our smart home or that the city council implements in our streets?

In this paper, we have chosen to analyze two reference examples of decentralized systems: Wikipedia and Bitcoin, to explore their governance systems and measure their current democratic status. Our aspiration is to use the found results as a reference for the future IoT models and apply necessary improvements to the eventual breaches that we have highlighted in both models.

In this context, the research questions that the paper aims to address are the following:

- *How can democracy be measured in Online Communities?*
- *To what extent do Bitcoin and Wikipedia communities respect this democracy?*
- *How can the study of Bitcoin and Wikipedia governance be used as a reference for governance of IoT decentralized systems?*

The methodology to answer these questions gives the following organization to the paper. In Section 2, we set our theoretical framework by successively defining Internet Governance, Governance Models in Online Creation Communities, and the pillars of Democratic Governance in Internet. Once our framework is defined, we discuss and evaluate Bitcoin Governance in Section 3. The Section is concluded by a table summarizing the level of compliance of Bitcoin to our proposed democracy pillars. A similar work is then done in Section 4 for Wikipedia. In Section 5, we extend our scope to IoT, by explaining how the results that we found for both models may be useful in decentralized IoT scenarios. In Section 6, we provide suggestions to improve the studied models, so that democracy principles are more respected. Finally, we conclude our paper in Section 7 by summarizing the findings and presenting future work in the area.

As observed, a theoretical approach was preferred to an empirical one for two main reasons. First, theory is the essence of this study that aims to encourage improving governance models. The scope regards the theoretical model itself and how it can be improved to be more democratic. Second, applying an empirical approach would limit the result of the study to the examples of Bitcoin and Wikipedia, while the aim is the opposite one; To use the examples as theoretical references for IoT systems governance. Future work would then be directed towards improving and building on the presented reference governance models instead of restarting another empirical study for each new system or community that is to be investigated.

## II. THEORETICAL FRAMEWORKS

### A. Internet Governance

The initial definition of Internet governance was suggested during the World Summit on the Information Society (WSIS) in Tunis, 2005, endorsed by the UN general assembly: "Internet Governance is the development and application by Government, the private sector, and civil society, in their respective roles, of shared principles, norms, rules, decision-making procedures, and programs that shape the evolution and use of the Internet" (UN DESA, 2005). This initial and broad definition limits the stakeholders to three selected parts: Governments, private sector and civil society. Now in 2019, new actors (international organizations, online communities) and ways of use have emerged, implying a necessity to update the previous definition. Kurbalija (2016) has listed the five groups within which the main Internet Governance issues fall: Infrastructure and standardization, Legal, Economic, Development and Sociocultural.

Since Internet Governance is a broad subject, our focus in this paper will be about the decision-making process in the so-called Online Creation Communities, defined in the section below.

### B. Governance Models in Online Creation Communities

Online Creation Communities (OCC) as defined by Morell (2010) are "networks of individuals that communicate, interact and collaborate; in several forms and degrees of participation which are eco-systemically integrated; mainly via a platform of participation on the Internet, on which they depend; and aiming at knowledge making and sharing".

In an OCC, as well as in more generic online projects, the two diametrically opposed governance types are: fully-centralized control by a single individual or organization (benevolent dictatorship) in opposition to distributed control awarded in recognition of contributions (meritocracy). We have chosen to present these models as they represent two opposite governance perspectives. In reality, it is hard to find them in such a level of clarity, as applied models in practice combine aspects from both.

- *Benevolent dictator governance model*

Introduced by Gardler and Hanganu (2013a), the benevolent dictator model (BDM) is steered by a unique leader, who leads the project, while the community manages it according to the strategical lines drawn by the leader. In case of disagreement, it is the benevolent dictator's responsibility to resolve disputes within the community and to ensure that the project can progress in a coordinated way. They have the last word. The community's role is to follow the decisions of the dictator and achieve the goals through active engagement and contribution.

The four types of participants in the benevolent dictator governance model are:

- ***Benevolent dictator.*** Leads the technical direction of the project and organizes the community. Has a final say on project decisions and community disputes. Not necessarily chosen (or agreed about) by the community, as it can be the person who started the project initially.
- ***Committers.*** Contributors who made several valuable contributions to the project and have a certain influence over the community. They have a power awarded to them by the leader, but only to be used following the strategies chosen by the dictator, not their own. Committers can either be chosen by the community or appointed by the dictator.
- ***Contributors.*** Community members participating in the project with any meaningful input, but not having the power of committers. Everybody can become a contributor. There is no specific skill requirement or selection process. They can be committers with experience and contributions.
- ***Users.*** Any member of the community having a need for the project. They are usually the majority. There is no specific requirement to be a user. Users motivated to develop the project can become contributors.

Nearly all the projects start with this centralized model, as they it has a limited number of contributors, and the decisions are made by this small team. Decision-making is in fact easier when centralized [4]. However, as a successful project grows, more and more people will be willing to contribute in the project. The decision-making process may need to evolve accordingly.

- *Meritocratic governance Model*

The second model that was also introduced by Gardler and Hanganu (2013b) is the Meritocratic governance model. In this model, the participants gain gradually influence over the project through the recognition of their contributions. Anyone with an interest in the project can join the community, contribute to the project design and participate in the decision-making process. The operating way is almost completely 'flat' structure, meaning that anyone willing to contribute can engage with their projects at any level.

The relevant roles that are suggested for this model are the following:

- ***Users.*** Similar role to the one described in the BDM. Any member of the community having a need for the project. They are usually the majority. There is no specific requirement to be a user. Users motivated to develop the project can become contributors.
- ***Contributors.*** Similar role to the one described in the BDM. Community members participating in the project with any meaningful input, but not having the power of committers. Everybody can become a contributor. There is no specific skill requirement or selection process. Users can become committers with experience and contributions.

- ***Committers.*** A committer is a contributor who has proven willingness and ability to participate in the project by providing valuable contributions over a period of time. A committer does not have authority nor makes decisions, but the rights and tools that they can access help them to follow the guidelines and rules drawn by the leadership.
- ***Project Management Committee (PMC).*** Members of this committee have additional responsibilities over the committers. They participate in the strategic planning and approve changes to the governance model. They make also decisions if a conflict is not resolved within the community. To become a PMC member, a committer must be nominated and voted by the other PMC members.

In its beginning, a meritocratic project will look very similar to a Benevolent dictator one, since the number of decision makers is small, and the process will be centralized. The key difference is that the starting team provides a transition mechanism for the distribution of control from the 'dictator' to a flat structure recognizing contribution, while in the BDM the dictator keeps their centralized power during the entire process.

*Comparison between the two models*

After presenting the two governance models above, we have summarized and compared their main characteristics in the Table 1 below.

TABLE I. SUMMARY AND COMPARISON BETWEEN BENEVOLENT DICTATOR AND MERITOCRATIC GOVERNANCE MODELS.

| | ***Benevolent Dictator*** | ***Meritocratic*** |
|---|---|---|
| **Overview** | - Benevolent dictator draws the strategic lines of the project.<br>- Community follows the dictator's guidelines and contributes in day-to-day work. | - Anyone interested and motivated can reach a position in the PMC, allowing to be part of decision-making.<br>- Community follows the PMC strategy. |
| **User** | No prerequisite to be a user. Majority in the community are "simple" users. | No prerequisite to be a user. Majority in the community are "simple" users. |
| **Contributor** | Contributors are active users who participate in the project. No skill is needed. | Contributors are active users who participate in the project. No skill is needed. |
| **Committer** | Most active contributors become committers, have access to more rights, but just to apply decisions from leadership. | Most active contributors become committers, have access to more rights, but just to apply decisions from leadership. |
| **Benevolent Dictator Vs PMC selection process** | Benevolent Dictator:<br>- Self-appointed.<br>- Usually the person who started the project.<br>- Remains during the entire process | PMC:<br>- Nominated and voted by PMC members.<br>- Members can be changed if needed (not active, not following rules, etc.) |
| **Contribution Process** | Anyone can contribute in the project, as long as the strategical line is followed. | Anyone can contribute in the project, and gradually take part in the decision-making process. |
| **Decision Making Process** | - The dictator makes the decisions.<br>- The dictator has the last word in case of a conflict in the community.<br>- The dictator informs the community about the decisions. | - PMC discuss and agree on decisions.<br>- PMC has the last word in case of a conflict in the community.<br>- Some decisions can be discussed within the community (through mailing list/forum for example). |
| **Conclusions** | - Centralized model.<br>- Opaque and unilateral decision-making process.<br>- Easy to manage.<br>- Most of power in the hands of a minority.<br>- Dictator is not accountable.<br>- Ultimate levels of power never accessible. | - Decentralized model.<br>- Transparent and discussed decision-making process.<br>- Complex to adopt.<br>- Power is given/removed by the community |

As a complement to this table, it is worth mentioning that it is possible for a project to combine both approaches and have a "hybrid" status. It is also possible for a project to make a transition from BDM to meritocratic model, given that there is a will from the benevolent dictator to share their power with the community. The opposite is certainly possible too, specifically when projects become very conflictual and the governance model is unclear.

It is also useful to mention that the two governance models above are valid for other platforms than OCC, such as online communities or network of peers in which no content is created but manage other goods.

### C. Democratic Governance in Internet

After defining Internet governance and going through the two main governance models in OCCs, the theoretical framework that we need to define for our research purpose is democratic governance in Internet. This will allow us to check later on if the examples that we are studying are complying with the "Internet democracy" metrics that we recommend.

In its broadest definition, modern democracy is a "system of government by the whole population or all eligible members of a state, typically through elected representatives" (Oxford Dictionaries, 2019). Democratic governance (also referred to by Good Governance) is the range of processes through which a society reaches consensus on and implements regulations, human rights, laws, policies and social structures – in pursuit of justice, welfare and environment protection (Cheema and Maguire, 2001). Good Governance and Accountability are principles of democracy governance models. They are about respecting human rights and the rule of law, wide and effective people's participation and transparent and accountable processes and institutions.

This political definition related to the "offline" concepts of states and societies can also be applied to online communities. In the Internet context, democratic governance is schematized through the multistakeholder approach, which defines the heart of the Internet ecosystem. The multistakeholder approach is the inclusion of the different stakeholders of the Internet ecosystem in the decision-making processes (Internet Society, 2016). This process reflects the commitment to an open dialog between governments, private sector organizations, civil society, and the technical community to shape the growth of the Internet and the policies that support and protect it.

The two different multistakeholder models existing nowadays are the "consultative model" and the "collaborative model" (GCSC, 2018).

In the **consultative model**, governments "consult" with non-governmental stakeholders, but the final decision remains in their hands. The risk is that non-governmental actors are just figurative players and their opinion is ignored during the decision-making. This model allows actors to raise their voice but does not guarantee that it is taken into account.

In the **collaborative model**, state and non-state actors are operating on an equal footing. The process of policy development is bottom-up, open and transparent. The outcome is rough consensus, where accountability is integrated into the checks and balances of the created mechanisms.

Given the definitions above, and regarding the OCC ecosystems that we will investigate, we have decided to appoint five principles as the online democracy principles that will be measured in the paper. We consider these principles as being the most relevant for our study. The principles are the following:

**Open participation** – The ability to participate for all in the online democratic process. It is a basic pillar in "classic" democracy (Klein et al, 2011). It is not only a right, but also a responsibility.

**Transparency** – Also a pillar of democracy (Gurría, 2011). It is awareness and ability to get information about what decisions are being made, by whom and why.

**Equality** – All participants in the project are valued equally, have equal opportunities, and are not discriminated.

**Accountability** – Those who are elected or appointed should be accountable for their actions (Behn, 2001; Hutchings, 2003). They must take decisions for the public interest.

**Democratic dispute management** - Conflict is inevitable in any system. Dispute management in a democratic way is as a necessary function of democratic governance.

## III. BITCOIN GOVERNANCE

### A. *Presentation of Bitcoin and Blockchain Technologies*

Bitcoin is a cryptocurrency that was introduced by Satoshi Nakamoto (nickname for an anonymous person or group) through a white paper written in 2008 (Nakamoto, 2008). In Bitcoin, electronic payments are performed by generating transactions directly between the users. Bitcoin runs a consensus protocol where any participant node that can solve a proof-of-work (POW) hash puzzle is allowed to add new blocks containing new transactions to the blockchain. The operation of solving the POW is called mining, and the users involved in it are miners.

A distributed consensus protocol in Bitcoin must satisfy: Agreement all honest nodes decide for the same value; Termination all honest nodes must terminate in finite time; and Validity a decision value must be the input value of an honest node (Coulouris et al., 2005). Bitcoin's consensus protocol is secure under the assumption that more than 50% of the participants are honest. Through this definition, we can easily notice that Bitcoin does not fall into the OCC grouping (since it is related to economics, not knowledge). Bitcoin is in fact an online network of peers, managing goods other than knowledge, but presents the same characteristics as an OCC in terms of governance model. We will then consider the two OCC governance models as applicable to Bitcoin in the remainder of the study.

Bitcoin has witnessed lately a wide attention worldwide (Reynolds, 2017). One of the reasons for this attention was the claim that it is a low-cost and decentralized digital currency that is independent of governments, central banks or any other centralized authority (Bitcoin Wiki, 2019).

Blockchain is the technology behind Bitcoin. The Blockchain is a chain of blocks archiving every Bitcoin transaction (Tapscott and Tapscott, 2016). Each block contains a number of transactions and is attached to the previous ones in the chain. Blockchain is typically managed by a peer-to-peer network collectively adhering to a protocol for validating new blocks. Once recorded, the data in any given block cannot be altered retroactively without the alteration of all subsequent blocks, which requires collusion of the network majority (blockchain GitHub, 2019). Currently, Bitcoin is governed by the consensus of a number of entities of different levels of influence, but Bitcoin governance is decided primarily by miners and the core development team.

### B. *Bitcoin Governance Model*

One of the advantages attracting users to Bitcoin is the governance model it claims to have. Bitcoin states that it is independent (not controlled by any entity or person) and decentralized. A decentralized governance is needed to ensure no single individual or entity can control the currency and take advantage of the system. This was the criticism of central banks and what gave birth to Bitcoin (Jamali and Pantoja, 2016). Decentralization means also that nobody has unilateral control. In this context, Bitcoin presents even a very particular characteristic: its creators are anonymous, and nobody knows who they are.

In its beginnings, Bitcoin followed the benevolent dictator model, where Satoshi Nakamoto set the strategical direction in the early development stage. The model shifted gradually to become a "consensus based" governance model. Bitcoin has a volunteer "core team" of developers who maintain the code and make updates as needed. They don't own bitcoin, but they act as a "control." Anyone can submit changes or suggestions to the community through GitHub (Bitcoin GitHub, 2019), these changes will be put up for vote (simple majority) and implemented by the core team when consensus is reached. The process of decision-making as well as conflict management follow both the consensus rule. It is however not rare that serious disagreements occur and cause a split in the community. It has already happened in Bitcoin, and users disagreeing with the core developers could fork the code and create their own currency (Azouvi et al., 2018).

The different stakeholder groups in the Bitcoin platform are presented in Table 2 below.

TABLE II. GROUPS OF STAKEHOLDERS, ROLES AND POWERS IN BITCOIN.

| *Stakeholder* | *Characteristics Roles* | *Incentives* | *Power* |
|---|---|---|---|
| **Developers (Core team)** | - Volunteer basis / Hired by the Bitcoin Foundation.<br>- Write the rules and the software.<br>- Manage improvement proposals.<br>- Manage conflicts.<br>- Encourage transparency through forums (GitHub). | - Privileged administrative rights.<br>- Social recognition. | - Decision-making power.<br>- Can block users even without consensus.<br>. |
| **Miners** | - Verify the transactions.<br>- Add transaction records to the Blockchain via PoW. | - Rewarded by Bitcoins.<br>- Seek maximizing profit by verifying a maximum of transactions. | - Voting power, proportional to their computational power.<br>- Can fork the blockchain if they don't share the same future view. |
| **Users** | - Participate in transactions.<br>- Send and receive Bitcoins.<br>- Do not influence the appointment of the core developers. | - Can push the Bitcoin price up or down.<br>- Peer-to-peer system not under the state authority. | - Suggestive power: Free to share opinions.<br>- Can decide on the Bitcoin price if they coordinate their way of use (boycott for example).<br>- Vote with their wallet: The more you pay, the more weight you have in the chain. |

From Table 2, we can observe that some aspects of meritocracy are present in the Bitcoin model, as anyone can be part of each of the three roles, depending on their knowledge and involvement in the community.

### C. *Evaluation of the Bitcoin Governance model*

While being secured and decentralized via a proof-of-work hash puzzle, Bitcoin's proof-of-work system has led it to become a centralized system as currently few miners have the privilege to add more blocks thanks to their huge investment in sophisticated and powerful hardware equipment to mine new bitcoins (Sedrati et al., 2017). The proof-of-work system is in fact estimated to require as much electricity as all of Denmark by 2020 (Deetman, 2016).

Several operations related to Bitcoin governance were investigated by (Gervais et al., 2014), concluding that the Bitcoin decision process is not completely decentralized as claimed. The main points that were criticized are the following:

- Mining became a centralized operation. More than 75% of Bitcoin computing power is controlled by 6 major mining pools.
- Less than 10 entities (supported by the core team) took a decision to outvote the majority of computing power in the network.
- The decision to vote (or not vote) for improvement proposals requested by the community is made unilaterally by the core team. The community can propose, but there is no guarantee that the proposition will go to vote, even if supported by the majority.

In parallel with the points noted in the paper above, we have investigated and found an additional number of noticeable issues and inconsistencies:

- In 2012, the Bitcoin foundation was created (Bitcoin Foundation, 2019). It is under the American law, which is in contradiction with the claim of independence from any state or entity.
- Due to the need of consensus, some conflicts can prove taking an enormous amount of time, or even not reach a solution. The debate on the block size limit is a good example (Clifford, 2017), remaining for years, before finally resulting on members of the community leaving Bitcoin.
- The Bitcoin governance is lacking guidance over responsibility. There is no explicit "legal responsible" in case a transaction is not executed correctly. Either collective responsibility or a special legal entity should be implemented.
- As presented in Table 2, miners are mostly working for their own interest (collecting Bitcoins). This can conflict with the needs of users or the good of the community. Miners do in fact prioritize their profit, and their votes go towards this point more than anything else.
- The core development team is not accountable for the changes and decisions that it implements. There is no clear process of punishment or revocation of power. In case of dispute, (lower-ranked) members will leave the community.

In addition to these issues, many censorship cases related to the Bitcoin community happened in the past. A summarizing list is presented by John Blocke (2016).

Taking into consideration the information gathered in this section, and given the democracy principles introduced earlier, Table 3 summarizes the correlation between Bitcoin governance and the five principles.

TABLE III. LEVEL OF RESPECT OF THE FIVE DEMOCRACY PRINCIPLES BY BITCOIN GOVERNANCE.

| *Democracy Principle* | *Respected in Bitcoin Governance?* |
|---|---|
| **Open Participation** | Yes |
| **Transparency – Regulation, Legislation** | Mostly transparent, lacking some legislation. |
| **Equality – On Rights and Responsibilities** | Partially, users with powerful computers make the system centralized. |
| **Accountability** | No |
| **Democratic Dispute Management** | No. A heavy dispute results either on a fork or a unilateral decision from the core team. |

## IV. WIKIPEDIA GOVERNANCE

### A. Presentation of Wikipedia and Wikimedia

Wikipedia is a multilingual collaborative free-content online encyclopedia based on a model of openly editable content. It was created for the mission of “providing free knowledge for all”. Since its creation in January 2001, Wikipedia has drastically grown. It is available in 303 languages (January 2019) and is one of the largest reference websites in the world. It is in fact ranked as the fifth most visited website worldwide (Alexa Internet Traffic, 2019). Both Wikipedia source code and content are open and free to be used by all. With more than 40 million articles in 303 languages, and over 35 million editors in English language alone (Wikipedian statistics, 2019), Wikipedia is one of the largest OCCs. Through its way of work, size and influence, Wikipedia is a good reference example to analyze as an Online Creation Community.

Wikipedia is only one project in the list of 15 projects that are under the supervision and support of the Wikimedia Foundation (WMF). WMF is a nonprofit organization supporting Wikipedia, with offices in San Francisco, USA. Its goal is to “keep the Wikimedia sites fast, secure, and available to all, defending Wikipedia and its volunteer editors from legal threats, building new features and tools to make it easy to read, edit, and share knowledge from Wikimedia sites, and by supporting the communities of editors who contribute to Wikipedia and the Wikimedia sites. WMF helps also bring new knowledge online, lowers barriers to access, and makes it easier for everyone to share what they know” (F.A.Q Wikimedia Foundation, 2019).

Wikipedia relies on the “Openly editable content” principle. This means that anyone can freely edit the encyclopedia (to different extents). There is in fact no restriction to be introduced to editing. All ages, cultures and backgrounds are encouraged to add or edit content. In Wikipedia, communities are defined by language, and not by national state for example. The reason is that each language Wikipedia is an independent encyclopedia. Users speaking and editing in a given language form the community. Some communities are therefore spread over many countries, while others could be all gathered in a given region of one country. However, decisions enforced by the WMF affect all the communities, since it is the transcendent entity managing the project in its totality.

On practical terms, making the content possible to be edited by anyone requires a significant number of rules, policies and roles. They are in fact necessary to prevent vandalism, protect the neutrality of the encyclopedia, solve disputes and manage the content.

The next section will detail the processes used in the governance of the encyclopedia, as well as the different roles involved in it. In our work, and since each local community sets its own rules, we have decided to restrict ourselves to the English Wikipedia (which is the biggest in term of numbers). Governance in other communities tend to approach the English-speaking model when growing, and this is the reason for choosing it as a reference. It is also worth to mention that strategic direction is set by the board and does not depend on a local community or language in particular.

### B. Wikipedia Governance Model

Similarly to the Bitcoin, Wikipedia started following the benevolent dictator model. It was created in 2001 by Jimmy Wales and Larry Sanger, who shaped its primary strategy. Many rules shaped by Wales and Sanger are still present in Wikipedia such as: respecting the other users, be neutral and not taking part in controversies. Attracted by the innovative idea, many people joined the movement and the community witnessed a big growth in a short time. The community dynamic has then started to shift the model towards a consensus-based model, with a voting system affecting several actions related to the encyclopedia. An extensive description of all the processes of Wikipedia Governance is provided at (Wikipedia: Governance, 2019).

From our perspective, we have divided governance in Wikipedia into two main parts: Global governance and local community governance.

**Global governance** regards the decision-making processes that are related to Wikipedia in general, as well as setting up the strategical direction. The Wikimedia foundation (WMF, founded in 2003) was created for that aim. Its main purpose is to support and manage the Wikimedia projects. WMF employs full-time employees in order to fulfill this role, some of them can be new to Wikipedia, while others can be members of the community who were hired in a certain position. Over the WMF there is the so-called “board of trustees” (Wikimedia: Board of Trustees, 2019). It has the ultimate authority of taking decisions related to Wikipedia. The board consists of ten trustees: One permanent seat for Jimmy Wales, five seats selected by the community (through different voting systems), and four seats appointed by the board itself on specific expertise needed.

On a **local scale**, communities organize themselves usually with regards to language. It is not rare however to have other gatherings, such as communities with a specific interest (photography, feminism, etc.). Our interest will be on the language communities, as the other communities are under their governance in each language Wikipedia. Since our reference is the English Wikipedia, we will briefly present the different roles involved in its local management, as well as their characteristics.

When a Wikipedia article is newly created, it follows usually the benevolent dictator model. The person who creates the article adds primary information and references to make it grow. Once it is known, other members of the community will start editing and improving it as well. It is however important to note that administrators can delete newly created articles for several reasons, such as lack of notability (Wikipedia: Notability, 2019). The main stakeholder groups involved in the editing process in Wikipedia are administrators, confirmed editors, and unregistered users.

There are also several offline roles related to the community (User Groups, chapters), but in this paper we have decided to highlight only the online governance. Information about offline affiliates and their operation mode can be found at Wikimedia movement affiliates list (2019).

Table 4 below summarizes the different roles that we think are relevant to understand the governance of Wikipedia, as well as their characteristics and power.

TABLE IV. GROUPS OF STAKEHOLDERS, ROLES AND POWERS IN WIKIPEDIA.

| ***Stakeholder*** | ***Characteristics Roles*** | ***Incentives*** | ***Power*** |
|---|---|---|---|
| **Board of Trustees** | - Appointed by the community / Hired for their expertise.<br>- Decide on the path to be followed by WMF. | - Highest decision power on Wikipedia.<br>- Social recognition. | - Decision-making power.<br>- High-level strategy decision for the future of Wikipedia.<br>- Power over the WMF.<br>- Set high level rules to be followed in all Wikimedia projects. |
| **Administrators** | - Manage the daily work on local Wikipedias.<br>- Appointed by the community given their engagement and dedication.<br>- Volunteers. They are not paid for the work. | - Highest decision power on the local community.<br>- Recognized as the "house-keepers" preventing vandalism. | - Can remove articles and users.<br>- Can grant or remove power to other users.<br>- Set rules and norms to be followed locally.<br>- Enforce decisions taken after conflicts.<br>- Can access and modify "fully-protected" pages. |
| **Confirmed editors** | - Appointed by administrators after showing dedication.<br>- Volunteers. They are not paid for the work. | - Participate in editing and preventing vandalism.<br>- Share the knowledge freely. | - Vote for the administrators, and indirectly for the Board of Trustees.<br>- Can edit with privilege. Their edits are not reviewed thoroughly, since they are "confirmed".<br>- Cannot remove article or users. |
| **Non-registered users** | - Access the encyclopedia and its content freely.<br>- Can edit if they want.<br>- Everyone can be a user, provided having access to a device. | - Read and share knowledge freely.<br>- Participate in building the biggest encyclopedia in the world. | - Can edit the articles if they want to, but their edits need to be reviewed in detail. |

### C. *Evaluation of the Wikipedia Governance Model*

The Wikipedia governance model is complex, and involves several actors, globally and locally, who work sometimes independently and separately from each other. The foundation of this governance system is decentralization, as well as maximization of consensus whenever it is possible. This independence from entities such as states is important to keep neutrality, one of Wikipedia pillars. One can refer to the episode that happened in 2017, when Wikipedia administrators refused to delete articles that Turkish officials wanted unilaterally to be removed. This has resulted in Wikipedia being blocked in Turkey (Roberts, 2017).

However, practice has shown that this model is not always decentralized and independent as it should. One evident consequence of the meritocratic model in Wikipedia is the loop that it creates. Most of the active users who become administrators and confirmed editors form a small "elite" who see the encyclopedia from their narrow angle. These experienced users tend to set rules from their own "advanced" perspective. This makes it very difficult for unaccustomed new users to be integrated in the system. A concrete example is that a new user creating their first article will certainly forget some step (or reference) causing their article to be immediately removed. A direct consequence is that the user might not want to continue editing, as the "unwelcoming" policies are in advantage of more experienced user, creating a "closed system".

In an interview conducted by one of the authors with Katherine Maher, the executive director of the WMF, a number of issues and challenges related to Wikipedia governance were discussed (interview with Katherine Maher, WikiIndaba, Tunis, 17 March 2018[1]). The main issues that were highlighted in the interview were:

- It is difficult to reach total equality for all users. One reason is that not all users have the same amount of free time to dedicate to the Wikipedia projects. A considerable number of hours is in fact needed in order to reach power roles in the encyclopedia. Given that it is a voluntary work, many users do not have the luxury of spending a lot of time in Wikipedia.
- The most important metric used to measure engagement in Wikipedia is the number of edits. This

[1] Link to the interview available on YouTube https://www.youtube.com/watch?v=zRU2xxv8o9I

metric is however subject of critics for many reasons. There is in fact no correlation between a user skilled in editing or adding content, and their ability to manage conflicts or set rules. It is not rare that users complain about the unfriendly behavior of many administrators, who reached their positions just with a significant number of edits, and few human and communication skills.

- Access to internet and devices is unequal in the world. A consequence of this situation is the centralization of knowledge. A majority of editors, articles and references are from the West, while knowledge about regions such as Africa is very limited. This situation is also reflected in a number of rules that are very "western"-centered. It is for instance difficult to accept oral sources, while they are in many cases the only sources in several cultures.
- A number of updates must be performed directly from the WMF, such as software updates, or technical updates in general (the visual editor is one known case). These updates are "forced" by the WMF over the communities (Miquel-Ribé, 2016), who can reject it on a later stage. There is a tendency to change this process by involving the communities earlier in discussions about their needs. However, this is a complex question as the software for Wikipedia is one, while each community can have their own opinion over a given feature.
- Accountability in smaller communities is difficult to control. It is in fact easy for the first comers in a given language to set their own rules and close the doors, making it a "mini-dictatorship". If the community is really limited and there are no complaints, it can take a very long time before such issues are noticed and solved.

Taking into consideration the information that we gathered in this section, and given the democracy principles that we introduced earlier, Table 5 summarizes the correlation between Wikipedia governance and the five principles.).

TABLE V. LEVEL OF RESPECT OF THE FIVE DEMOCRACY PRINCIPLES BY WIKIPEDIA GOVERNANCE.

| *Democracy Principle* | *Respected in Wikipedia Governance?* |
|---|---|
| **Open Participation** | Yes |
| **Transparency – Regulation, Legislation** | Yes |
| **Equality – On Rights and Responsibilities** | Mostly. Users with more time and easier access are more likely to have decision-making roles in the future. |
| **Accountability** | Mostly. Not all the board members are voted.<br>In small communities, admins may not be accountable if no control. |
| **Democratic Dispute Management** | Mostly. Some decisions can be "forced" by the WMF if the community is split. |

## V. DECENTRALIZED GOVERNANCE IN IOT CONTEXTS

Internet of Things is considered as the natural evolution for the future of Internet. The concept of independent connected objects will change several aspects and technologies that are seen as trivial and relevant in the present. In this section, we will present two fictional IoT scenarios in order to familiarize with the concept. The first part shows a direct link between these scenarios and the use of Bitcoin and Wikipedia. In the second part, we have chosen a more abstract approach by explaining how the governance of Wikipedia and Bitcoin can be a reference to IoT scenarios, even without a direct link. Finally, the last part will give some concrete suggestions in order to improve these governance systems, taking into consideration real-life scenarios.

### *A. IoT context and scenarios*

By definition, IoT implies that Internet will not only be restricted to specific devices aimed to be connected (smartphones, laptops, etc.). With IoT, most of "objects" need to be connected, besides their primary role. In this part, we will present two IoT related scenarios that use Wikipedia and Bitcoin. By presenting these scenarios, the idea is to demonstrate how Wikipedia and Bitcoin governance will be influential, not only for these platforms and their communities, but also for many IoT scenarios that will be directly linked to them and affect our daily lives.

Several studies and researches have been investigating potential applications of Blockchain within Internet of Things. One of the main challenges facing this implementation is scalability (Sedrati et al., 2017). In fact, the current blockchain is not scalable enough to be used on a large extent. A number of initiatives and projects are working to address this problem by reparametrizing metrics to improve throughput and lower latency. An interesting example to be read is the paper "*On scaling decentralized blockchains*" (Croman et al., 2016). The Lightening network (LN) layer, proposed by Poon and Dryja (2016) is also envisioned to help solving scalability challenges by enabling instant micropayments and lower payment fees.

Moreover, transactions are not fast enough to be used in all contexts, and double-spending attacks can still occur (Karame et al., 2015). Taking these limitations into account, the scenarios that we are presenting in this part are to be considered on a lower scale and are not likely to be subject to the attacks presented above.

- **Scenario 1: Smart City – Tourism in the Medina**

In a smart city, tourists will be able to find digital information related to the monuments they are visiting on the monuments themselves. Initiatives such as TarfayaPedia[2] and MedinaPedia[3] are already allowing this to happen. Tourists walking in the respective Medinas (old towns) can scan QR codes found on different monuments and will be redirected to their Wikipedia page in their own language. Building on this concept, a system providing suggestions of items to purchase can be implemented. Souvenirs and relevant objects can be bought using bitcoins and shipped directly to the tourists' home country.

[2] TarfayaPedia project https://en.wikipedia.org/wiki/Wikipedia:GLAM/TarfayaPedia

[3] MedinaPedia project https://meta.wikimedia.org/wiki/Wikimedia_Tunisie/MedinaPedia

- **Scenario 2: Smart House – Fridge in control**

In a smart house, most of objects if not all will be connected, either directly with each other or online, or both. In this configuration, simple scenarios can be imagined and implemented in order to simplify different tasks of daily life. One example of such scenarios is the functionalities that a fridge will fulfill when being "smart". When a smart fridge will be lacking items necessary for the household, it can order them digitally itself and pay the transaction using bitcoin. The owner can also provide a list of meals to be prepared during the working week. After looking for the necessary ingredients, the fridge can order and purchase them. During this process, and given the different preferences (diet, allergy etc.) the fridge can check compatibilities in Wikipedia and report any potential risk. Wikipedia can also be used to help calculate the nutritional value of the meal, in order to provide a balanced nutrition. heads.

### B. *Applications of Bitcoin and Wikipedia governance in IoT scenarios*

Many IoT systems have a decentralized architecture, allowing objects to be connected to each other without the need of a central authority. In this context, the results of the present paper can be a reference, as Wikipedia and Bitcoin governance models may be applied to similar structures. Two examples of relevant IoT scenarios are presented below.

- **Scenario 1: Smart City – Connecting Services**

The concept of Smart city has several definitions and is subject to different interpretations depending on the context and the area. The reference paper from Hollands (2008) presents the challenges in defining this concept, and the various meanings that it can imply. Besides this ambiguity, we believe that all the different definitions have an aspect in common: In a smart city, a number of services will be connected with each other, and their governance can be disputed from different parts with various levels of responsibility. Examples of these parts are ministry of tourism (monuments, guiding information in the city), ministry of transportation (red light, smart signs) and the city council (parking, museums). Each of the smart objects under the responsibility of one part will "directly" be connected to another "object" falling into the responsibility of another actor. It is necessary to have several agreements and to synchronize this decentralized model. In this context, governance models of Wikipedia and Bitcoin can be a starting point to manage a decentralized smart-city.

- **Smart House: Connecting Home Appliance**

Similarly to a smart city, a smart house has a number of services connected to each other (fridge, oven, heating system, TV, etc.). An important difference is however that the actors are different. The house owner can choose to apply a benevolent dictator model over the system and take unilateral decisions about the system. It can also be possible to configure the system so that it makes its own decisions with time. Technologies such as artificial intelligence (AI) can be applied in smart houses (Augusto and Nugent, 2006). In this sense, governance models that we have analyzed in this paper can be helpful to develop AI governance algorithms in a smart house.

## VI. Suggestions for improvements

In our work, we have investigated the governance models of both Wikipedia and Bitcoin, then we have stressed their importance and explained why they can be considered as references for future IoT decentralized systems. In both models, we have also noticed several points that can be improved. Our suggestions to tackle some are the following:

- **Enhance decentralization.** Using decentralized mining pools as suggested in by Gervais et al. (2014) is an example of improvements that can be made to "retrieve" decentralization in Bitcoin. Another improvement was suggested by Paul et al. (2014), through modifying the block header to introduce fair competition between the miners.

  Building on these ideas, a guideline to be followed can be the constant control of the mining pools/admin list in order to avoid any "looping" that may centralize power. This control can be done by an independent entity, such as the board of trustees in Wikimedia, or a new board that will be created for this purpose in the Bitcoin. Control can have different purposes, such as verification of the respect of democracy principles in "small Wikipedias" that can have long lasting dictatorships. It can also be in form of setting metrics at different points of time to consult and review the "decentralization" status.

  Not only control of the current status is needed, but also active actions that encourage new participants to join the community and thus make the decentralization continuous. The Wikimedia foundation is already working on this point (interview with Katherine Maher), but further efforts are required. Bitcoin from its side does not have any active strategy to involve new members in the community.

- **More transparent decision making.** We have observed in our study that not all decisions can be traced by the community. In some cases, information about different decisions exist but are difficult to find. It is understandable that informing all the community about all events and decisions might not be necessary, but it can be interesting to discuss with the community itself about the decisions that are most important to be informed about. The current situation is that decision makers decide unilaterally (perhaps with good intentions) about what are the decisions that are most relevant to inform, how they were taken, and which channel to use to update the community. Consensus within community about decision making process is important for its well-being. It will also allow flexibility and dynamism, as priorities might change with time, and interests as well.

- **Ensure single vote per person or entity.** A problem encountered within the decentralized model is that anonymity allows one single individual to process several accounts. Being active enough, this person can pass the voting criteria with the different accounts (having at least 50 edits in the English Wikipedia for instance). Obviously, this is a non-democratic process as it breaches the equality principle. One way used to counter this is to verify the IP address of the sender. This measure is however not enough as the voting process is usually long (more than a week) and the user might connect from different locations and devices. A measure that we suggest is to link users with their electronic IDs. In fact, countries such as Sweden (BankID, 2019) and Italy (Carta Di Identità Elettronica, 2019) have their own national electronic ID, used to identify the person online for different purposes. In a future IoT context, it is possible to use this identification system on all "things" owned by a person, including in this case online accounts in Wikipedia and Bitcoin. Not only this will prevent single users from voting more than once, but it will also ensure that voters are human, and not programmed bots for example. One should mention however that privacy questions should be investigated before implementing this solution.

- **Investigate new decentralized models.** The early adoption and relative success of Bitcoin does not necessarily imply that its governance model is the most democratic. Different blockchain-based solutions have different models, which can present consequent improvements and variations from the Bitcoin model. One interesting cryptocurrency model is Dash (2014). It is a decentralized autonomous organization (DOA) that was forked from the Bitcoin software in 2014. The main novelty brought by this system is the introduction of master nodes (computers running a Dash wallet and making decisions) as intermediaries in the transactions. Dash developers receive payments for their contributions, which might encourage new users and increase decentralization.

  A second example is the Aragon project (2019), an open source decentralized application based on blockchain, that allows anyone to create and manage a decentralized organization.

## VII. Conclusions and future work

Governance of Internet of Things requires a careful investigation of the different existing models, in order to improve and adapt them for future use. Our aim with this paper is to participate in this effort, by studying both Bitcoin and Wikipedia models as references for decentralized systems. After defining our online democratic framework and its metrics, we have evaluated the compliance of Bitcoin and Wikipedia current governance models towards these metrics. Our analysis showed that governance within both Bitcoin and Wikipedia does not respect democracy principles in different extents. Despite claims of decentralization, Bitcoin community remains largely centralized and power rests in hands of few miners. Accountability and dispute management are also areas to be improved for this cryptocurrency. Wikipedia does better in comparison, but its scattered nature creates accountability problems in smaller language communities, where few users can do what they want. The intervention of Bitcoin and Wikimedia Foundation in these communities weakens also the democratic process, as these foundations operate under the American jurisdiction, and cannot guarantee a decentralized community that does not depend on any state or entity.

Through this paper, we hope that Bitcoin and Wikipedia communities will take note of our findings in order to foster decentralization and ensure the respect of democracy principles in their models.

By pointing out the limitations of Bitcoin and Wikipedia models, we hope that future research will be shading light on improvements and new alternatives for decentralized systems. We believe that the governance models developed for future IoT systems will be inspired from reference communities similar to Bitcoin and Wikipedia. In section 5, we have linked between IoT scenarios and the reference examples, in order to highlight the importance and relevance of our findings for the future of IoT governance. In section 6, we have presented suggestions to improve the current governance models. It is important to note that there is no universal democratic governance model that can be applied to all contexts. A simple difference could be the comparison between IoT systems at home as opposed to those in public environments. Principles such as open participation, transparency, equality, accountability and democratic dispute management might have different meanings depending on the context, as private discussions at home follow usually a different process than an international agreement between sovereign countries. Moreover, the scope of our study was highlighting active users. Future studies can involve passive ones. Not only logged in passive users should be considered as such, but it can also be extended to mailing list readers or participants in discussion spaces.

The models that we have presented and assessed present the advantage of being flexible and can be adapted to different situations, even if the debates will be naturally distinct and will involve diverse stakeholders.

With the current work, we wish that future research in IoT governance will be directed towards enhancing decentralization, more transparent decision making, and ensuring equality. Once these pillars are settled, IoT systems will find a stronger democratic background that can be implemented to govern the Internet of the future.

## References

Alexa Internet Traffic Statistics for Wikipedia.org. [online] https://www.alexa.com/siteinfo/wikipedia.org (Accessed January 20th, 2019)

Aragon Project. [online] https://aragon.org/ (Accessed January 20th, 2019)

Augusto, J. C., & Nugent, C. D. (Eds.). (2006). Designing smart homes: the role of artificial intelligence (Vol. 4008). Springer.

Azouvi, S., Maller, M., & Meiklejohn, S. (2018, March). Egalitarian Society or Benevolent Dictatorship: The State of Cryptocurrency Governance. In 22nd International Conference on Financial Cryptography and Data Security.

Bank ID, Sweden. [online] https://www.bankid.com/en/ (Accessed January 20th, 2019)

Behn, R. D. (2001). Rethinking democratic accountability. Brookings Institution Press.

Bitcoin Documentation on GitHub. [online] https://github.com/bitcoin/bips (Accessed January 20th, 2019)

Bitcoin Foundation. [online] https://bitcoinfoundation.org/ (Accessed January 20th, 2019)

Bitcoin Wiki. [online] https://en.bitcoin.it/wiki/ (Accessed January 20th, 2019)

Blockchain Documentation on GitHub. [online] https://github.com/itext/i7j-pdfchain (Accessed January 20th, 2019)

Blocke J. (2016) A (brief and incomplete) history of censorship in /r/Bitcoin. Medium. [online] https://medium.com/@johnblocke/a-brief-and-incomplete-history-of-censorship-in-r-bitcoin-c85a290fe43 (Accessed January 20th, 2019)

Carta Di Identità Elettronica. Ministry of Interior, Italy. [online] https://www.cartaidentita.interno.gov.it/ (Accessed January 20th, 2019)

Cheema, G. S., & Maguire, L. (2001). Governance for human development: the role of external partners. Public Administration and Development, 21(3), 201–209.

Clifford J. (2017) Understanding the Block Size Debate. Medium. [online] https://medium.com/@jcliff/understanding-the-block-size-debate-351bdbaaa38 (Accessed January 20th, 2019)

Coulouris, G. F., Dollimore, J., & Kindberg, T. (2005). Distributed systems: concepts and design. pearson education.

Croman, K., Decker, C., Eyal, I., Gencer, A. E., Juels, A., Kosba, A., ... & Song, D. (2016, February). On scaling decentralized blockchains. In International Conference on Financial Cryptography and Data Security (pp. 106-125). Springer, Berlin, Heidelberg.

Dash Cryptocurrency. [online] https://www.dash.org/ (Accessed January 20th, 2019)

Deetman, S. (2016). Bitcoin could consume as much electricity as Denmark by 2020. Opinion Article, Leiden University. https://motherboard.vice.com/en_us/article/aek3za/bitcoin-could-consume-as-much-electricity-as-denmark-by-2020 (Accessed January 20th, 2019)

English Oxford Dictionaries – Democracy definition. [online] https://en.oxforddictionaries.com/definition/democracy (Accessed January 20th, 2019)

Fuster Morell, M. (2010). Governance of online creation communities: Provision of infrastructure for the building of digital commons (Doctoral dissertation).

Gardler, R., & Hanganu, G. (2013). Benevolent dictator governance model. OSS Watch, 7. http://oss-watch.ac.uk/resources/benevolentdictatorgovernancemodel (Accessed January 20th, 2019)

Gardler, R., & Hanganu, G. (2013). Meritocratic governance model. OSS Watch, 7. http://oss-watch.ac.uk/resources/meritocraticgovernancemodel (Accessed January 20th, 2019)

Gervais, A., Karame, G., Capkun, S., & Capkun, V. (2014). Is bitcoin a decentralized currency? IEEE security & privacy, 12(3), 54-60.

Global Commission on the Stability of Cyberspace (GCSC) (2018). Towards A Holistic Approach for Internet Related Public Policy Making. https://cyberstability.org/wp-content/uploads/2018/02/GCSC_Kleinwachter-Thought-Piece-2018-1.pdf (Accessed January 20th, 2019)

Governance of Wikipedia. English Wikipedia Page. [online] https://en.wikipedia.org/wiki/Wikipedia:Governance (Accessed January 20th, 2019)

Gurría, A. (2011). Openness and Transparency-Pillars for Democracy, Trust and Progress. Presented on occasion of the launching of the Open Government Partnership. http://www.oecd.org/unitedstates/opennessandtransparency-pillarsfordemocracytrustandprogress.htm (Accessed January 20th, 2019)

Hollands, R. G. (2008). Will the real smart city please stand up? Intelligent, progressive or entrepreneurial? City, 12(3), 303–320.

Hutchings, V. L. (2003). Public opinion and democratic accountability: How citizens learn about politics. Princeton University Press.

Internet Society (2016). Internet Governance – Why the Multistakeholder Approach Works. https://www.internetsociety.org/resources/doc/2016/internet-governance-why-the-multistakeholder-approach-works/ (Accessed January 20th, 2019)

Jamali, R., Li S., Pantoja R. (2016) Digital Asset Class of the Future – Bitcoin vs Ethereum. The Economist. [online] https://www.economist.com/sites/default/files/economist_case_comp_ivey.pdf (Accessed January 20th, 2019)

Karame, G. O., Androulaki, E., Roeschlin, M., Gervais, A., & Čapkun, S. (2015). Misbehavior in bitcoin: A study of double-spending and accountability. ACM Transactions on Information and System Security (TISSEC), 18(1), 2.

Klein, A., Kiranda, Y., & Bafaki, R. (2011). Concepts and Principles of Democratic Governance and Accountability. A Guide for Educators. Uganda, Kampala: KAS Publishers. http://www.kas.de/wf/doc/kas_29779-1522-2-30.pdf?11121919022 (Accessed January 20th, 2019)

Kurbalija, J. (2016). An introduction to internet governance. Diplo Foundation.

Miquel-Ribé, M. M. (2016). Identity-based motivation in digital engagement: the influence of community and cultural identity on participation in wikipedia (Doctoral dissertation, Universitat Pompeu Fabra).

Nakamoto, S. (2008). Bitcoin: A peer-to-peer electronic cash system.

Notability definition. English Wikipedia Page. [online] https://en.wikipedia.org/wiki/Wikipedia:Notability (Accessed January 20th, 2019)

Paul, G., Sarkar, P., & Mukherjee, S. (2014, December). Towards a more democratic mining in bitcoins. In International Conference on Information Systems Security (pp. 185-203). Springer, Cham.

Poon, J., & Dryja, T. (2016). The bitcoin lightning network: Scalable off-chain instant payments. [online] https://lightning.network/lightning-network-paper.pdf (Accessed January 20th, 2019)

Reynolds, M. (2017). Psychology explains why your friends can't shut up about bitcoin. Wired. [online] http://www.wired.co.uk/article/will-bitcoin-bubble-burst-or-crash-2017-psychology-cryptocurrency-economics (Accessed January 20th, 2019)

Roberts, R. (2017, April). Turkey blocks Wikipedia over 'refusal to delete articles alleging government terrorist links'. The Independent. [online] https://www.independent.co.uk/news/world/turkey-blocks-wikipedia-refusal-to-delete-articles-terrorism-co-operation-erdogan-media-censorship-a7710491.html (Accessed January 20th, 2019)

Sedrati, A., Abdelraheem, M. A., & Raza, S. (2017). Blockchain and IoT: Mind the Gap. In Interoperability, Safety and Security in IoT (pp. 113-122). Springer, Cham.

Tapscott, D., & Tapscott, A. (2016). Blockchain revolution: how the technology behind bitcoin is changing money, business, and the world. Penguin.

United nations Department of Social and Economic Affairs (UN DESA) – Definition of Internet Governance. https://publicadministration.un.org/en/internetgovernance (Accessed January 20th, 2019)

Wikimedia Foundation. Board of Trustees. [online] https://meta.wikimedia.org/wiki/Wikimedia_Foundation_Board_of_Trustees (Accessed January 20th, 2019)

Wikimedia Foundation. Frequently asked questions. [online] https://foundation.wikimedia.org/wiki/FAQ/en (Accessed January 20th, 2019)

Wikimedia movement affiliates list. [online] https://meta.wikimedia.org/wiki/Wikimedia_movement_affiliates/Portal (Accessed January 20th, 2019)

Wikipedian Statistics, Number of editors. [online] https://en.wikipedia.org/wiki/Wikipedia:Wikipedians#Number_of_editors (Accessed January 20th, 2019) 1955. *(references)*